\documentclass[a4paper,12pt]{article}

\usepackage[english]{babel}
\usepackage[utf8x]{inputenc}
\usepackage[T1]{fontenc}

\usepackage[a4paper,top=3cm,bottom=2cm,left=3cm,right=3cm,marginparwidth=2cm]{geometry}

\usepackage{setspace}
\usepackage{amssymb}

\usepackage{cite}

\usepackage{amsmath}
\usepackage{graphicx}
\usepackage[colorinlistoftodos]{todonotes}
\usepackage[colorlinks=true, allcolors=blue]{hyperref}

\title{Active Tuning of Resonant Lattice Kerker Effect}

\author{Lei Xiong$^{1,2,3}$, Hongwei Ding$^{1,*}$, Yuanfu Lu$^{2,3,4}$, and Guangyuan Li$^{2,3,4,*}$}

\date{}
\begin{document}
\maketitle

\begin{spacing}{2.0}

\noindent $^1$School of Information Science and Engineering, Yunnan University, Kunming 650500, China

\noindent  $^2$CAS Key Laboratory of Human-Machine Intelligence-Synergy Systems, Shenzhen Institute of Advanced Technology, Chinese Academy of Sciences, Shenzhen 518055, China

\noindent $^3$Guangdong-Hong Kong-Macao Joint Laboratory of Human-Machine Intelligence-Synergy Systems, Chinese Academy of Sciences, Shenzhen Institute of Advanced Technology, Shenzhen 518055, China

\noindent $^4$Shenzhen College of Advanced Technology, University of Chinese Academy of Sciences, Shenzhen 518055, China


\noindent *Corresponding authors: dhw1964@163.com; gy.li@siat.ac.cn

\end{spacing}

\newpage

\begin{abstract}
The Kerker effect has been generalized in nanophotonics and meta-optics, and has recently been of great interest by relating to various fascinating functionalities such as scattering management and perfect transmission, reflection or absorption. One of the most interesting generalizations is the resonant lattice Kerker effect in periodic nanostructures. However, its active tuning has not been explored yet. Here, we report, for the first time, the active control of the resonant lattice Kerker effect in periodic Ge$_2$Se$_2$Te$_5$ nanodisks. By changing the crystalline fraction, we show that the electric dipole lattice resonance (ED-LR), the magnetic dipole resonance (MDR), and thus the resonant lattice Kerker effect are all red-shifted. We therefore realize the transition from the ED-LR to the resonant lattice Kerker effect, which enables multilevel tuning of reflection, transmission and absorption with modulation depths above 86\%. Taking advantage of the MDR redshifts, we also observe broadband and multilevel tuning of transmission with modulation depth of 87\% over a broadband range of 588~nm. Our work establishes a new path for designing high-performance active nanophotonic devices.
\end{abstract}

\newpage

\section{Introduction}
The Kerker effect refers to the totally eliminated backward or forward scattering phenomenon discovered by Kerker {\sl et al.} in 1983 \cite{JOSA1983Kerker}. With the rapid development of meta-optics, the Kerker effect has been generalized and can be realized by spectrally overlapping the electric dipole resonance (EDR) and the magnetic dipole resonance (MDR), or by overlapping multipoles of different natures and orders \cite{OE2018Yuri_KerkerRev,PRL2019_GKerker}. This unique effect has attracted increasing interest and has found a variety of intriguing applications, ranging from scattering control including anti-reflection and anti-transmission \cite{ACSNano2013_zeroBW,JOSAB2017Babicheva_antiR,OE2018_zeroFWBW,NanoP2020Guo_Kerker2ARAT,OE2021Magnusson_Kerker}, to Huygens' sources \cite{AOM2015Yuri_KerkerHuygens,PRM2020KerkerHuygens} and the generalized Brewster effect \cite{NC2016_GBrewster,OE2018_GBrewster}.

Recently, Babicheva and Evlyukhin proposed the concept of {\sl resonant lattice Kerker effect} in two-dimensional periodic nanoparticle arrays with specially chosen periodicity \cite{LPR2017Babicheva_LatticeKerker}. They showed that by varying one period of the rectangular lattice, the electric dipole lattice resonance (ED-LR) can be excited and spectrally overlaps with the MDR of single nanoparticles in the vicinity of Rayleigh anomalies (RAs), leading to strong suppression of reflection or backward scattering from the array as a whole. Based on the similar principle, they also reported resonant suppression of transmission or forward scattering \cite{OL2018Babicheva_LatticeKerkerT}. Since ED and MD lattice resonances can be independently controlled by varying the mutually perpendicular array periods, these lattice resonances can be brought into spectral overlap without imposing limitations on particle shape \cite{NanoP2018Babicheva_LatticeKerkerSiRod}. As a result, the resonant lattice Kerker effect was also observed in periodic silicon nanodisk arrays \cite{NanoP2018Babicheva_LatticeKerkerSiRod,JAP2018_LatticeKerkerSiRod}, and more recently in plasmonic nanostructures \cite{PRB2021Karpov_LatticeKerkerPlas,JPCC2021_LatticeKerkerPlas,JOSAB2021Karpov_LatticeKerkerPlas}. For a lossless system, suppressed reflection corresponds to perfect transmission \cite{NanoP2018Babicheva_LatticeKerkerSiRod,JAP2018_LatticeKerkerSiRod}, whereas suppressed transmission corresponds to absolute 100\% reflection \cite{OME2020_LatticeKerkerPerfectRef}. When moderate loss is involved, however, the forward scattering can be significantly suppressed together with the backward scattering, resulting in greatly enhanced absorption or even perfect absorption \cite{ACSP2018_LatticeKerkerPerfectAbs, OL2021_LatticeKerkerPerfectAbs,OE2020_LatticeKerkerPerfectAbs}. Although it has been shown that the polarization and the angle of incidence can provide two additional degrees of freedom to tune these lattice resonances \cite{MRSC2018Babicheva_LatticeKerkerOblique,P2020Ilia_LatticeKerkerOblique}, the active tuning of the resonant lattice Kerker effect has not been explored yet. 

In the field on active nanophotonics, phase-change materials such as Ge$_2$Sb$_2$Te$_5$ (GST) have been widely adopted owing to their unique merits. These include high refractive index, dramatic optical contrast between amorphous and crystalline states, non-volatile, rapid and reversible switching characteristics, relatively low loss, and high chemical and long-term stability \cite{NatP2017Taubner_PCMreview,AOM2019Bozhevolnyi_GSTreview,ATS2019Cao_PCMreview,NanoP2020Adibi_PCMreview}. Based on GST nanodisk metasurfaces, broadband switching between the EDR and the MDR \cite{OE2018Qiu_GSTEDMDvsR}, or between the EDR and the anapole state \cite{NC2019Bozhevolnyi_GSTanapole} was proposed or demonstrated, showing extinction modulation between 10\% and 60\%. Making use of the dynamic spectral shift of the EDR in Si/GST/Si nanodisks, multilevel reflection modulation between 10\% and 80\% was also demonstrated \cite{Optica2020Wright_SiGSTSi}. Employing other phase-change materials such as GeTe or Ge$_2$Sb$_2$Se$_4$Te (GSST), the switching between the Kerker (zero backward scattering) and anti-Kerker (vanishing forward scattering) regimes \cite{ACSP2019_Non2SupScatPCM}, or the quasi-continuous reflection modulation between 10\% and 40\% \cite{NN2021_EtuneGST} was reported. In these metasurfaces, however, lattice resonances are not involved and the active tuning is restricted to Mie resonances only. Quite recently, some of the authors adopted GST in the form of films functioning as surrounding media for metallic nanostructures and realized dynamical tuning of plasmonic surface lattice resonances  \cite{NanoMat2021Shi_SLRGSTthick,RiP2021Shi_SLRGSTthin}.

In this work, we report the active tuning of the ED-LR, the MDR, and thus the resonant lattice Kerker effect in GST metasurfaces. We observe, for the first time, the transition from the ED-LR to the resonant lattice Kerker effect by changing the GST crystallinity. This leads to multilevel tuning of reflection over a strikingly large range between 94.1\% and 1.4\% by exploiting the intermediate phases of GST. This modulation depth is much larger than those based on Mie resonances of single nanoparticles \cite{OE2018Qiu_GSTEDMDvsR,NC2019Bozhevolnyi_GSTanapole,Optica2020Wright_SiGSTSi}, highlighting the advantages of tuning the Mie lattice resonances. We also observe continuous shifting of the MDR, and the associated modulation of transmission with modulation depth of 87\% over a broadband range of 588~nm. Hence, the GST metasurfaces can function as an actively reconfigurable optical switch with high modulation depths and multilevel control abilities. 

\section{Design and simulation setup}

\begin{figure*}[tbh]
\centering
\includegraphics[width=0.95\linewidth]{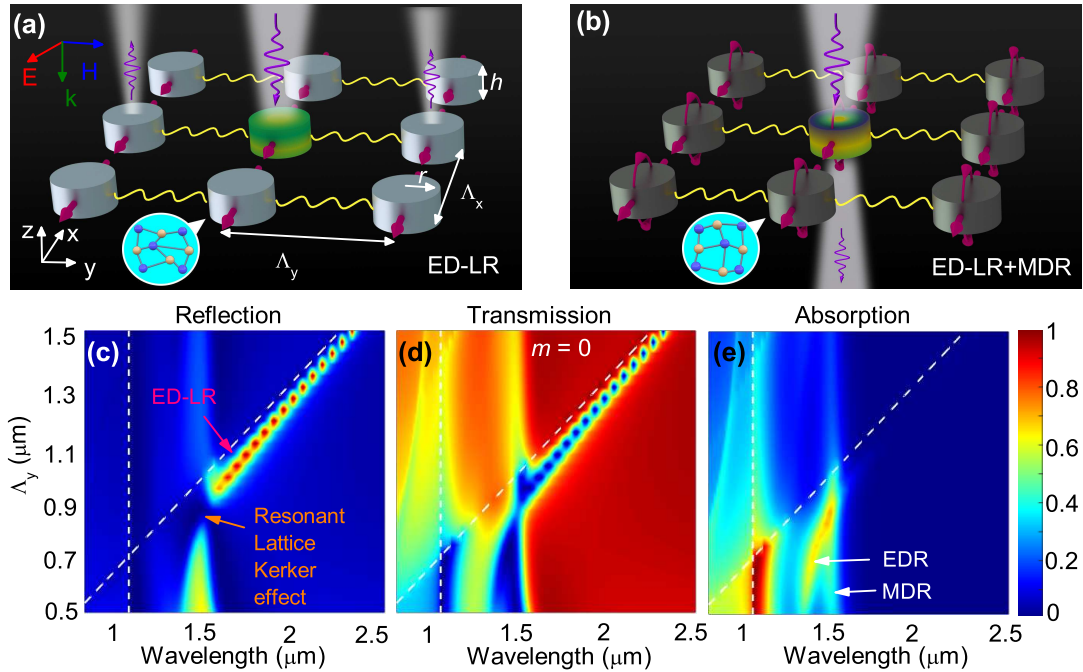}
\caption{(a)(b) Schematics of a GST metasurface composed of periodic nanodisks. The metasurface supports (a) ED-LR in amorphous phase and (b) resonant lattice Kerker effect in crystalline phase, leading to strong and active tuning of reflection and transmission. (c) Reflection, (d) transmission and (e) absorption versus wavelength and period $\Lambda_y$ for metasurfaces in amorphous state. The white lines denote RAs: $\lambda_{{\rm RA},(1,0)}=1.05~\mu$m (vertical dashed) and $\lambda_{{\rm RA},(0,1)}=n_0\Lambda_y$ (oblique dash-dot).}
\label{fig:schem}
\end{figure*}

Figure \ref{fig:schem}(a)(b) illustrates the ED-LR and the resonant lattice Kerker effect supported by the 2D array of periodic GST nanodisks in amorphous and semi-crystalline states, respectively. The metasurface is embedded in homogeneous surrounding medium with refractive index $n_0=1.5$, and is normally illuminated by a plane wave linearly polarized along the $x$-axis. Each nanodisk has radius $r=200$~nm and height $h=220$~nm such that it supports EDR around 1.136~$\mu$m and MDR around 1.535~$\mu$m in the amorphous state according to the Mie theory calculations. These GST nanodisks are periodically patterned with the period in the $x$ direction fixed to be $\Lambda_x=0.7~\mu$m, and varying period in the $y$ direction, $\Lambda_y$.

The reflection and transmission spectra, as well as the near-field distributions of the silver/gold nanodisk array were simulated using an in-house package for fully vectorial rigorous coupled-wave analysis (RCWA), which was developed following \cite{JOSAA1995RCWA,JOSAA1997RCWA,PRB2006RCWA}. In all the simulations, we adopted retained orders of $41\times41$, which were shown to be large enough to reach the convergence regime. 

The effective dielectric constants of Ge$_2$Sb$_2$Te$_5$ in different crystallization levels can be described by the Lorenz--Lorentz relation \cite{AJP1982LLmodel},
\begin{eqnarray}
\label{Eq:epsiEff}
\frac{\varepsilon_{\rm eff}(\lambda)-1}{\varepsilon_{\rm eff}(\lambda)+2}=m\frac{\varepsilon _{\rm c}(\lambda)-1}{\varepsilon_{\rm c}(\lambda)+2}+\left ( 1-m \right )\frac{\varepsilon_{\rm a}(\lambda)-1}{\varepsilon_{\rm a}(\lambda)+2} \,.
\end{eqnarray}
Here $m$, ranging from 0 to 1.0, is the GST crystalline fraction, and $\varepsilon_{\rm a}(\lambda)$ and $\varepsilon _{\rm c}(\lambda)$ are the permittivities of amorphous ($m=0$) and crystalline ($m=1.0$) GST at wavelength $\lambda$, respectively. The values of $\varepsilon_{\rm a}(\lambda)$ and $\varepsilon _{\rm c}(\lambda)$ were taken from \cite{ProcSPIE2017GSTnk}.

\section{RESULTS AND DISCUSSION}
\subsection{Resonant lattice Kerker effect in GST metasurfaces}

When $\Lambda_y$ varies from 0.5~$\mu$m to 1.5~$\mu$m, the EDR is excited around $\lambda=1.34~\mu$m for small $\Lambda_y$, and the MDR, which is independent of $\Lambda_y$, is excited around $\lambda=1.499~\mu$m. These wavelengths are close to their counterparts of individual nanodisks. For period $\Lambda_y$ and wavelength positions that are close to the $(0,\pm{1})$ order RA line, for which $\lambda_{{\rm RA},(0,\pm{1})}=n_0 \Lambda_y$, the ED-LR is excited instead. In this scenario, the reflection approaches unity, and both the transmission and the absorption are vanishing, as shown by figure~\ref{fig:schem}(c)--(e). Hence, the metasurface now acts as a perfect mirror. These findings are consistent with the 2D arrays of silicon nanospheres \cite{LPR2017Babicheva_LatticeKerker} or nanodisks \cite{NanoP2018Babicheva_LatticeKerkerSiRod}.

Particularly, when $\Lambda_y=0.86~\mu$m the ED-LR spectrally overlaps with the MDR at the wavelength of $\lambda_0=1.448~\mu$m, the resonant lattice Kerker effect takes place, as evidenced by the totally suppressed reflection. This corresponds to increased transmission and enhanced absorption. Note that the absorption stems from the intrinsic loss of the amorphous GST nanodisks in the near-infrared regime. 

According to the Mie theory, the spectral positions of the EDR and the MDR of single nanodisks heavily depend on the refractive index of the nanodisk. Therefore, we expect to observe redshifts (or blueshifts) of these Mie resonances by increasing (or decreasing) the GST refractive index. This can be done by changing the GST crystalline phase via thermal, electrical or optical stimuli \cite{NatP2017Taubner_PCMreview}. The effects on the Mie lattice modes and the resonant lattice Kerker effect, however, have not been investigated yet.

\subsection{Active tuning of resonant lattice Kerker effect}

\begin{figure*}[tbh]
\centering
\includegraphics[width=0.95\linewidth]{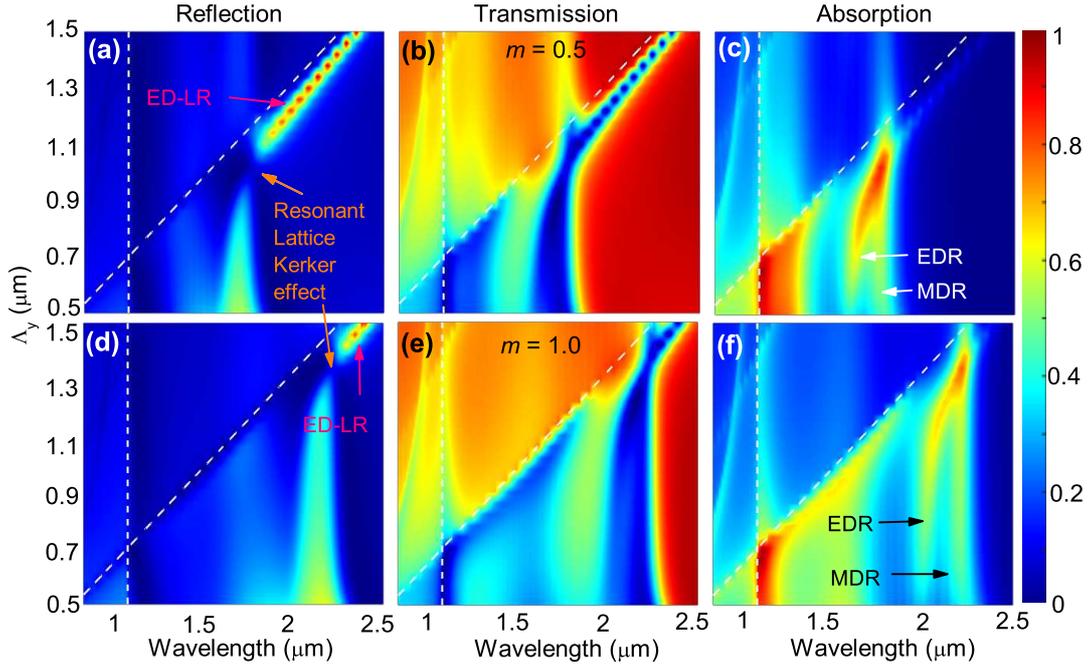}
\caption{Similar to Figure~\ref{fig:schem}(c)--(e) but for (a)--(c) semi-crystalline ($m=0.5$) and (d)--(f) crystalline ($m=1.0$) phases.}
\label{fig:RTA25_50}
\end{figure*}

We now investigate the effects of the crystalline phase on the optical properties of the GST metasurfaces. Figure~\ref{fig:RTA25_50} shows that as the crystalline fraction increases to $m=0.5$ or 1.0, the EDR, the MDR, and the ED-LR are all red-shifted owing to the increased refractive index. Hence, the wavelength of the resonant lattice Kerker effect is also red-shifted to $\lambda_{5}=1.679~\mu$m for $m=0.5$, or $\lambda_{10}=2.114~\mu$m for $m=1.0$. Comparing the spectra in figures~\ref{fig:schem}(c)--(e) and \ref{fig:RTA25_50}, we find that the ED-LRs are excited at both $\lambda_{5}$ and $\lambda_{10}$ when $m=0$, and at $\lambda_{10}$ when $m=0.5$. Similar phenomena can also be found for other $m$ values. Therefore, we observe the transition from the ED-LR to the resonant lattice Kerker effect when the GST crystalline fraction increases until reaching a certain value. This transition arises because the red-shifted MDR spectrally overlaps with the ED-LR, resulting in the resonant lattice Kerker effect at the original position of the ED-LR.

\begin{figure}[!hbt]
\centering
\includegraphics[width=0.5\linewidth]{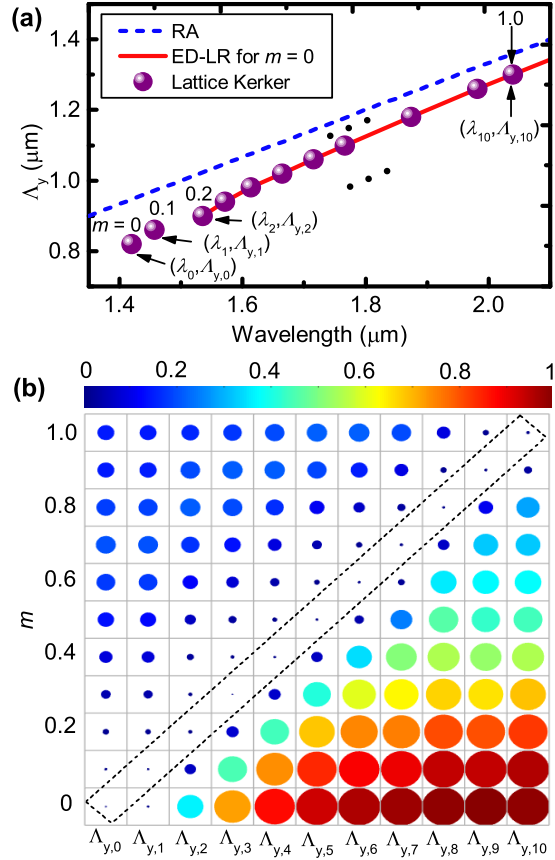}
\caption{(a) Period and wavelength positions of the ED-LRs for $m=0$ (red curve), and of the resonant lattice Kerker effect for different crystalline fractions (purple balls): $(\lambda_0,\Lambda_{y,0})$ for $m=0$, $(\lambda_1,\Lambda_{y,1})$ for $m=0.1$, ..., and $(\lambda_{10},\Lambda_{y,10})$ for $m=1.0$. The blue line indicates the $(0,\pm{1})$ order RA line. (b) Reflection versus period $\Lambda_{y,j}$ with $j=0,1,...,10$ and crystalline fraction $m$. The size and color of circles indicate the intensity of reflection. The dashed box encloses the $m$ values when the resonant lattice Kerker effect takes place.}
\label{fig:PvsWv_RvsmWv}
\end{figure}

We extract the period $\Lambda_y$ and wavelength positions of the ED-LRs for the GST metasurface in amorphous phase ($m=0$; figures~\ref{fig:schem}(c)--(e)), and plot these positions in a red curve in figure~\ref{fig:PvsWv_RvsmWv}(a). We also extract the period $\Lambda_y$ and wavelength positions of the resonant lattice Kerker effect for different crystalline fractions ranging from 0 to 1.0 (figures~\ref{fig:schem}(c)--(e) and \ref{fig:RTA25_50}), and plot these in purple balls: $(\lambda_0,\Lambda_{y,0})$ for $m=0$, $(\lambda_1,\Lambda_{y,1})$ for $m=0.1$, ..., and $(\lambda_{10},\Lambda_{y,10})$ for $m=1.0$. Strikingly, we find that the purple balls for $m\geq 0.2$ all locate on the red curve, which is almost parallel to the $(0,\pm{1})$ order RA line with a nearly constant spectral separation of 130~nm. This suggests the transition from the ED-LR when $m=0$ into the resonant lattice Kerker effect as $m$ increases to a certain value above 0.2.

Figure~\ref{fig:PvsWv_RvsmWv}(b) depicts the evolution of reflection for metasurfaces with period $\Lambda_{y,j}$ and operating at wavelengths $\lambda_{j}$ as a function of the crystalline fraction. Results show that, for $j=0$ or 1, the reflection approaches zero when $m=0$ or 0.1 due to the resonant lattice Kerker effect. As $m$ increases, the reflection first gradually increases to $\sim 20\%$ and then decreases slightly. For $j\geq2$, on the contrary, the reflection is initially high for small values of $m$ owing to the excitation of the ED-LR. As $m$ increases, the reflection first gradually decreases to near zero because of the resonant lattice Kerker effect, and then slightly increases. Interestingly, the specific value of $m$ required for achieving the resonant lattice Kerker effect continuously increases with $\Lambda_y$, as shown by the dashed box in figure~\ref{fig:PvsWv_RvsmWv}(b).

\begin{figure*}[!hbt]
\centering
\includegraphics[width=\linewidth]{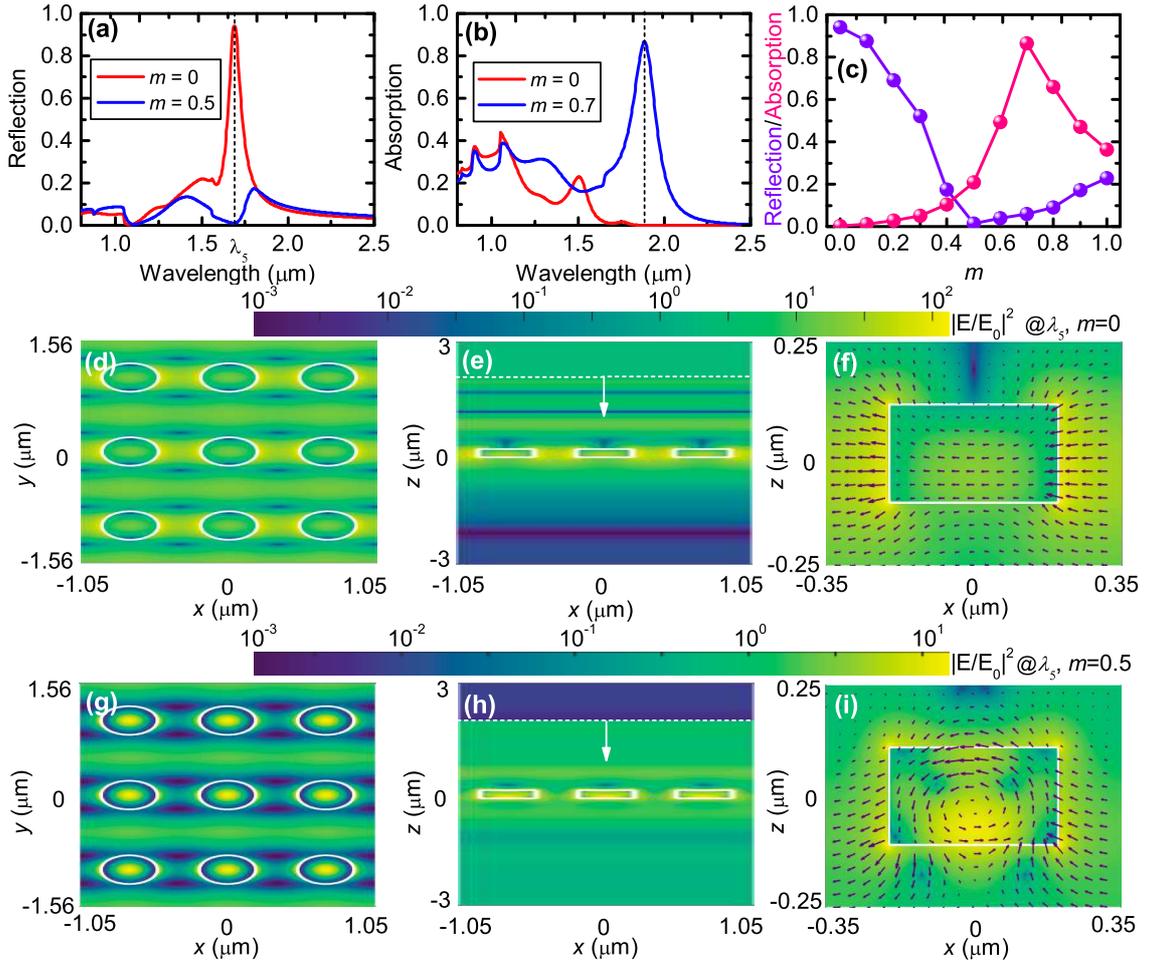}
\caption{(a) Reflection and (b) absorption spectra of the GST metasurface with period $\Lambda_{y,5}=1.04~\mu$m and two different crystalline fractions. At $\lambda_5=1.686~\mu$m the reflection reaches 94.1\% when $m=0$ and approaches zero when $m=0.5$, owing to the ED-LR and the resonant lattice Kerker effect, respectively. At $1.866~\mu$m the absorption is negligible when $m=0$ and reaches 86.6\% when $m=0.7$. (c) Reflection at 1.686~$\mu$m and absorption at $1.866~\mu$m versus $m$. (d)--(i) Near-field electric field distributions (color for intensity and arrows for directions) at $\lambda_{5}$ for (d)--(f) the ED-LR when $m=0$ and  (g)--(i) the resonant lattice Kerker effect when $m=0.5$. The white boxes outline the GST nanodisks, and the horizontal dashed lines in (e)(h) indicate the incident plane. }
\label{fig:RAvsWvmField}
\end{figure*}

We further exemplify the active modulation of reflection by a typical metasurface with $\Lambda_{y,5}=1.04~\mu$m and three crystalline fractions. Figure~\ref{fig:RAvsWvmField}(a) shows that at $\lambda_{5}=1.686~\mu$m the reflection reaches the peak value of 94.1\%  when $m=0$, but is vanishing (1.4\%) when $m=0.5$. These correspond to the ED-LR and the resonant lattice Kerker effect, respectively. Figure~\ref{fig:RAvsWvmField}(c) shows that as $m$ increases from 0 to 0.5, the reflection decreases continuously from 94.1\% to 1.4\% because of the transition from the ED-LR to the resonant lattice Kerker effect. As $m$ further increases, the reflection increases slightly to 22.9\%. Hence, our results demonstrate dynamically reconfigurable and multilevel control of reflection by exploiting the intermediate GST phases. Strikingly, the absolute modulation depth in reflection, which is defined as $M \equiv |R_{m'}-R_{m}|$, reaches as high as 93\%. This makes the proposed GST metasurfaces superior to those without the lattice resonances in the literature \cite{OE2018Qiu_GSTEDMDvsR,NC2019Bozhevolnyi_GSTanapole,Optica2020Wright_SiGSTSi}, highlighting the unprecedented benefits of the lattice resonances.

\begin{figure*}[!hbt]
\centering
\includegraphics[width=0.7\linewidth]{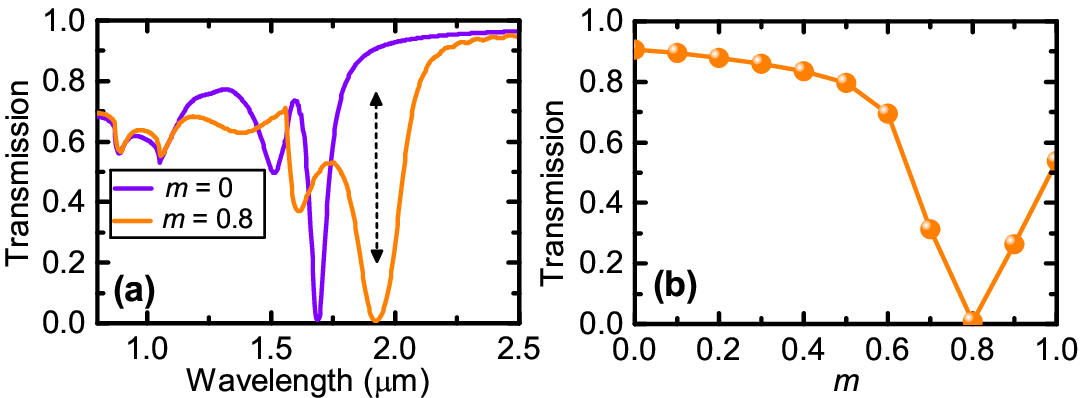}
\caption{(a) Transmission spectra of the GST metasurface with period $\Lambda_{y,5}=1.04~\mu$m. As $m$ increases from 0 to 0.8, both the EDR and the MDR are redshifted, resulting in switching of transmission from $T=90.6\%$ to 0.7\% at the wavelength of 1.922~$\mu$m. This corresponds to the modulation depth of 89.9\%. (b) Transmission at 1.922~$\mu$m versus $m$, showing multilevel tunability by exploiting intermediate crystalline phases.
}
\label{fig:TvsWvm}
\end{figure*}

On the other hand, high (or suppressed) reflection corresponds to weak (or enhanced) absorption and transmission. Figure~\ref{fig:RAvsWvmField}(b) shows that at 1.886~$\mu$m the absorption is negligible (only 0.4\%) when $m=0$, but reaches the peak value of 86.6\% when $m=0.7$. As $m$ increases, the absorption first increases from 0.4\% to 86.6\%, and then decreases to 36.2\% when $m=1.0$, as shown by figure~\ref{fig:RAvsWvmField}(c). These behaviors suggest that the absorption in the GST metasurface can also be actively modulated with multilevel tunability. Similarly, the transmission can also be actively controlled with large modulation depth of 89\% and multilevel tunability, as shown in figure~\ref{fig:TvsWvm}. However, this requires large change in crystallinity ($\Delta m=0.8$). The tuning of transmission requiring much smaller change in crystallinity will be discussed later.

In order to understand the physics behind the far-field reflection and absorption spectral features associated with the ED-LR and the resonant lattice Kerker effect, a near-field optical picture is required. Figure~\ref{fig:RAvsWvmField}(d)--(f) presents the simulated electric field distributions of the metasurface with period $\Lambda_{y,5}=1.04~\mu$m and $m=0$. It is clear that the electric field, which is aligned with the incident field and is greatly enhanced (${\rm max}\left\{|E/E_0|^2\right\}=135$), shows a lattice electric dipolar field pattern, and that the direction of the diffraction waves propagation is orthogonal to the electric field polarization. These are typical features of the ED-LR. The lattice electric dipolar moment generates considerable dipolar radiations, giving rise to near-unity reflection, as shown by figure~\ref{fig:RAvsWvmField}(e). On the other hand, because only a small portion of electric energy is confined in the GST nanodisks, and the intrinsic loss of amorphous GST is relatively low in the near-infrared regime, the absorption is very weak.

When $m=0.5$, figure~\ref{fig:RAvsWvmField}(g)--(i) shows that, besides the lattice electric dipolar fields that are confined outside the nanodisks, there exist strong and circulating electric fields that are confined in the nanodisks. Since the circulation of the electric fields
results in strong magnetic fields inside the nanodisk, the MDR is excited together with the ED-LR. The destructive interference between the MDR and the ED-LR results in the resonant lattice Kerker effect, as evidenced by the suppressed backward scattering (figure~\ref{fig:RAvsWvmField}(h)). Because a large portion of electric energy is confined in the GST nanodisks, and GST with larger crystalline fraction suffers from higher intrinsic loss, the absorption is enhanced when the resonant lattice Kerker effect takes place, as shown by figure~\ref{fig:PvsWv_RvsmWv}(c).

\subsection{Multilevel and broadband tuning of transmission}

\begin{figure*}[!tbh]
\centering
\includegraphics[width=\linewidth]{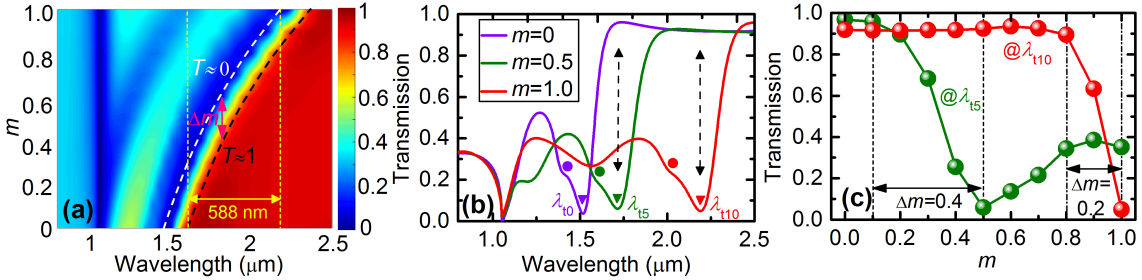}
\caption{(a) Transmission of GST metasurface with $\Lambda_y=0.5~\mu$m versus wavelength and crystalline fraction. The positions of $T\approx 0$ and the upper boundary of the $T\approx 1$ region are respectively denoted by the white and black dashed curves with a separation of $\Delta m\approx 0.3$ for eye guidance. (b) Transmission spectra for three different crystalline fractions of $m=0$, 0.5 and 1.0. Wavelengths of EDR are indicated by circles, while those of MDR are by triangles: $\lambda_{{\rm t}0}$, $\lambda_{{\rm t}5}$ and $\lambda_{{\rm t}10}$ for $m=0$, 0.5 and 1.0, respectively. (c) Transmission at $\lambda_{{\rm t}5}$ and at $\lambda_{{\rm t}10}$ versus crystalline fraction. The change in cystallinity required for switching from $T\approx 1$ to $T\approx 0$ is only $\Delta m=0.4$ for $\lambda_{{\rm t}5}$, or 0.2 for $\lambda_{{\rm t}10}$.}
\label{fig:TVsm}
\end{figure*}

While until now we considered the active tuning of reflection, absorption, and transmission taking advantage of the transition from the ED-LR to the resonant lattice Kerker effect in the vicinity of the $(0,\pm1)$ order RA, we find that the tuning of transmission due to the spectral shift of the MDR when $\Lambda_y$ and the wavelength are away from the RA is also appealing. In figures~\ref{fig:schem}(d) and \ref{fig:RTA25_50}(b)(e), we find that the transmission is near zero owing to the excitation of the MDR and approaches unity for wavelengths above the MDR. As $m$ increases, the transmission switches from $T\approx 100\%$ to $T\approx 0$ because of the redshift of the MDR. This is better visualized in figure~\ref{fig:TVsm}, where a typical GST metasurface with period $\Lambda_y=0.5~\mu$m is exemplified.

Figure~\ref{fig:TVsm}(a) shows that the MDR related to vanishing transmission undergoes continuous redshift as the crystalline fraction increases, as indicated by the white dashed curve. Accordingly, the smallest wavelength of the near-unity transmission region also increases. Therefore, for a given operation wavelength above 1.481~$\mu$m, the effective switching from $T\approx 100\%$ to $T\approx 0$ can take place within a small increase of the GST crystallinity, $\Delta m\approx 0.3$. Strikingly, this switching effect can be observed over an ultra-broadband range over $\Delta \lambda=588$~nm (from $1.604~\mu$m to $2.192~\mu$m), as indicated by the vertical dotted lines.

Figure~\ref{fig:TVsm}(b) shows the transmission spectra for three typical crystalline fractions. It is clear that both the EDR (denoted as circles) and MDR (triangles) are redshifted for larger $m$. Taking advantage of the MDR redshifts, we observe the  switching of transmission from 96.6\% to 6.0\% at the MDR wavelength of $\lambda_{{\rm t}5}=1.718~\mu$m if $m$ increases from 0 to 0.5, or from 92.2\% to 4.9\% at $\lambda_{{\rm t}10}=2.192~\mu$m if $m$ increases from 0.5 to 1.0. Accordingly, the absolute modulation depth in transmission reaches as high as 91\%, or 87\%, respectively.

By exploiting intermediate crystalline states within the small ranges of $\Delta m$, which are required for achieving the effective switching of transmission, we further demonstrate the multilevel tunability of transmission modulation in figure~\ref{fig:TVsm}(c). At the MDR wavelength of $\lambda_{{\rm t}5}$ (or $\lambda_{{\rm t}10}$), multilevel modulation of transmission can be realized by increasing the crystalline fraction from 0.1 to 0.5 (or from 0.8 to 1.0).

\section{CONCLUSIONS}

In conclusion, we have investigated the active tuning of the resonant lattice Kerker effect in metasurfaces composed of periodic GST nanodisks. We have shown that when the ED-LR is excited in the vicinity of the RA, the reflection reaches as high as 94\%, making the GST metasurface in amorphous state a nearly perfect mirror. By spectrally overlapping the ED-LR with the MDR through specially choosing the period $\Lambda_y$, we have realized the resonant lattice Kerker effect, as evidenced by the strongly suppressed reflection. We have shown that the EDR, the MDR, the ED-LR, and thus the resonant lattice Kerker effect are all red-shifted by increasing the GST crystallinity. Remarkably, we have observed the transition from the ED-LR to the resonant lattice Kerker effect by increasing $m$ until a specific value. This leads to multilevel tuning of reflection, transmission and absorption with modulation depths above 86\%. Taking advantage of the MDR redshifts, we have also observed broadband and multilevel tuning of transmission with large modulation depth of 87\% over a broadband range of 588~nm. Our results have further revealed that the active tuning of reflection and transmission requires a small change in crystallinity ($\Delta m \approx 0.3$), enabling low-intensity dynamic control. We expect that the absorption loss can be significantly reduced by designing GST metasurfaces in the mid-infrared regime \cite{NatP2017Taubner_PCMreview,AOM2019Bozhevolnyi_GSTreview}, or by adopting other low-loss phase-change chalcogenides such as GSST \cite{NC2019_GSST}. Therefore, this work opens the opportunity to design active nanophotonic devices with high performance and multilevel tunability.


\section*{Acknowledgments}
This work was supported by the State Key Laboratory of Advanced Optical Communication Systems and Networks, China (2020GZKF004).

\bibliographystyle{unsrt}
\bibliography{sample}

\end{document}